\documentclass[twocolumn]{aastex63}
\usepackage{amsmath,amstext}

\newcommand{\CfA}{\affiliation{Center for Astrophysics \textbar{} Harvard \& Smithsonian, 60 Garden Street, Cambridge, MA 02138-1516, USA}}

\newcommand{\Harvard}{\affiliation{Harvard College, Cambridge, MA 02138, USA}}

\begin{document}

\title{Tracing a Multi-Temperature Quiescent Prominence's Thermodynamic Evolution from Sun to Earth}

\author[0009-0001-1144-6345]{Callie A. García} \Harvard \CfA
\author[0000-0002-8748-2123]{Yeimy J. Rivera}\CfA
\author[0000-0002-6145-436X]{Samuel T. Badman}\CfA
\author[0000-0002-7868-1622]{John C. Raymond}\CfA
\author[0000-0002-6903-6832]{Katharine K. Reeves}\CfA
\author[0000-0001-6692-9187]{Tatiana Niembro}\CfA
\author[0000-0002-5699-090X]{Kristoff W. Paulson}\CfA
\author[0000-0002-7728-0085]{Michael L. Stevens}\CfA

\begin{abstract}
Solar prominences are cool, dense stable structures routinely observed in the corona. Prominences are often ejected from the Sun via coronal mass ejections (CMEs). However, they are rarely detected in a cool, low-ionized state within CMEs measured in situ, making their evolution hard to study. We examine the thermodynamic evolution of one of these rare cases where a quiescent prominence eruption clearly preserves its low-ionized charge state as evidenced by in situ detection. We use multi-viewpoint Extreme Ultraviolet (EUV) observations to track and estimate the density, temperature and speed of the prominence as it erupts. We observe that part of the prominence remains in absorption well beyond initial liftoff, indicating the bulk of the prominence experiences minimal ionization and suggesting any strong heating is balanced by radiative losses, expansion, or conduction. From its subsequent in situ passage near 1au, charge states reveal that the prominence is composed of both cool, low-ionized ions as well as hotter plasma reflected by the presence of highly ionized iron, Fe$^{16+}$. Simulated non-equilibrium ionization and recombination results using observationally derived initial conditions match the in situ multi-thermal state for a prominence composed of 70\% cool plasma with a 1.8MK peak temperature, and 30\% hot plasma with a 4.3MK peak temperature. This suggests that the prominence may not be heated uniformly or that parts of it cools more rapidly. The complex, multi-thermal nature of this erupting prominence emphasizes the need for more comprehensive spectral observations of the global corona. 
\end{abstract}

\section{Introduction}\label{sec:intro} 

Coronal mass ejections (CMEs) are driven by large scale reconnection events at the Sun which release massive amounts of plasma and magnetic field flux. CMEs are complex structures, seen in white light coronagraphs to contain many coherent features \citep{Illing1985}. Commonly, the brightest feature in white light is the core of a CME that is likely associated with a solar prominence (also known as a filament when viewed on the disk). Prominences at the Sun are cool ($\sim 10^{4}$ K), dense ($10^{9-11}$ cm$^{-3}$) features that appear magnetically suspended in the corona. A comprehensive review of prominences can be found in \cite{Parenti2014, Gibson2018}. 

When observed on the limb, prominences are often observed within a 'cavity' or more aptly named, hot prominence shrouds that surround it \citep{Habbal2010_hotshrouds}. The hot prominence shrouds are highly twisted magnetic environments, likely giving rise to the flux rope structure routinely observed within CMEs measured in situ \citep{Habbal2014}. Despite being immediately surrounded by a million Kelvin temperature solar atmosphere, the central core of prominences maintain much lower temperatures than their surroundings. As such, prominences are regularly seen in absorption in extreme ultraviolet (EUV) images as they are cool enough to contain neutral material. However, during the early stages of the eruption, part or most of the prominence can often be seen transitioning from absorption to emission, indicating rapid ionization and strong heating \citep{Landi2010, Wraback2024_multi_therm_prom}. In particular, it is found that out of all the elements making up a CME, the prominence experiences the largest amount of heating during the eruptive phase \citep{Murphy2011, Sheoran2023}. 

Solar filaments are often divided into one of two types: quiescent filaments are found in quiet regions of the Sun often at higher latitudes, and are referred to as polar crown filaments, while active filaments are situated within active regions, often near sunspots. Although quiescent filaments are longer lived at the Sun than active filaments, often observed for several solar rotations, both can become unstable and drive a CME. 

Occasionally, these eruptions are Earth-directed, driving global geomagnetic storms that generate aurorae. The physics behind how prominence material and the associated magnetic structure is ejected and evolves through interplanetary space is intimately connected to its impact on Earth, see detailed reviews from \citep{Webb2012LRSP, Riley2018SSR, Al_Haddad2025SSR}. To understand dynamics of the eruption, it is important to understand its early stages of evolution where magnetic energy is quickly transformed into thermal and kinetic energy \citep{Murphy2011, Linker2016, Torok2018, Lee2017, Reeves2019, Wilson2022}. The energy deposited to heat the prominence is one of the leading terms of the prominence's energy budget and often exceeds its kinetic energy in the corona \citep{Akmal2001}. Therefore, better constraints to the plasma's thermodynamic evolution can place specific heating and kinetic energy requirements important to improving propagation timescales and forecasting capabilities of CMEs. 

Overall, observations of prominences that maintain a low-charge state beyond the corona are difficult to observe. There have been a few cool prominence captured traveling within CMEs during eclipses and by space-based instruments \citep{Schmahl1977, Ciaravella1997, Ciaravella2000, Akmal2001, Raymond2002, Raymond2004, Landi2010, Lee_Raymond_2012, Ding2017}. Most recently, observations in the 304\AA~channel from the Extreme Ultraviolet Imager/Full Sun Imager telescope on Solar Orbiter indicate the presence of cool prominence material within a CME out to $>6R_{\odot}$, or solar radii \citep{Mierla2022} that was later observed in situ \citep{Palmerio2024}. However, although CME/prominence morphology are routinely observed by imagers, coupled spectral observations and in situ coverage of the events are infrequent. 

In the heliosphere, prominences observed in situ in a low-ionized state are rare ($\sim 4\%$; \citealt{Lepri2010}) as compared to the number of prominences embedded in CMEs seen leaving the corona (up to 70\%; \citealt{Gopalswamy2010, Giordano2013}). The prominence event we study is of particular interest as it is one of only 14 known to have been detected in-situ by heavy ion compositional detectors at L1 between $1997-2013$ \citep{Lepri2021}. In addition, it is also well-observed at the Sun. The event has also been examined as part of a series of interacting CMEs happening during this period \citep{Li2011, Harrison2012, Mostl2012}, however no detailed heavy ion analysis has been carried out. 

Our work investigates this quiescent prominence eruption at the Sun through multi-perspective remote images as well as in situ observations during its passage at the first Lagrange point (L1). The remote observations are used as sensitive constraints to its thermodynamic evolution just after liftoff where some of the largest acceleration and heating is observed, while the in situ bulk parameters constrain the final state of the prominence. 

Radial evolution of the fractional abundances are generated by a non-equilibrium ionization (NEI) code where the constrained thermodynamic profiles are optimized to reconstruct the full measured charge state distribution of carbon, oxygen, and iron of the prominence material measured in situ. 

Prior studies of CMEs model the ion abundances have used such NEI calculations to quantify the heating experienced as reflected in the measured ion abundances \citep{Rakowski2007, Gruesbeck2012, Rivera2019a, Laming2023}. The reason ions are excellent diagnostics of the plasma's thermodynamic evolution is that their in situ fractional abundances reflect the heating and cooling during these early stages of outflow as they ionize and recombine up to a freeze-in altitude. The freeze-in altitude marks the distances from the Sun where the ionization and recombination becomes ineffective as the density becomes too low. The freeze-in altitudes of the solar wind and CMEs span 1 to $>6$ R$_\odot$, significantly overlapping with the middle corona \citep[1.5-6 R$_\odot$][]{West2023}. Several studies have utilized heavy ion signatures to connect transients and the solar wind from their coronal origin to their heliospheric passage in situ \citep{Parenti2021, Yardley2024, Rivera2025}.

This paper is organized as follows. In Section \ref{sec:observations} we discuss observations of the event, both in situ and at L1. In Section \ref{sec:methods} we detail our process for determining near-Sun temperature, density, and velocity constraints, which we then apply to a non-equilibrium ionization simulation. In Section \ref{sec:results} we discuss the obtained constraints, and the plasma properties that most closely aligned the non-equilibrium ionization simulation to the measured fractional abundances of the plasma at L1. In Section \ref{sec:discussion} we evaluate the implication of these findings.

\section{Observations of the Event}\label{sec:obs}
\subsection{Remote Observations}
\label{sec:observations}

\begin{figure}[h!]
\centerline{\includegraphics[width=0.47\textwidth]{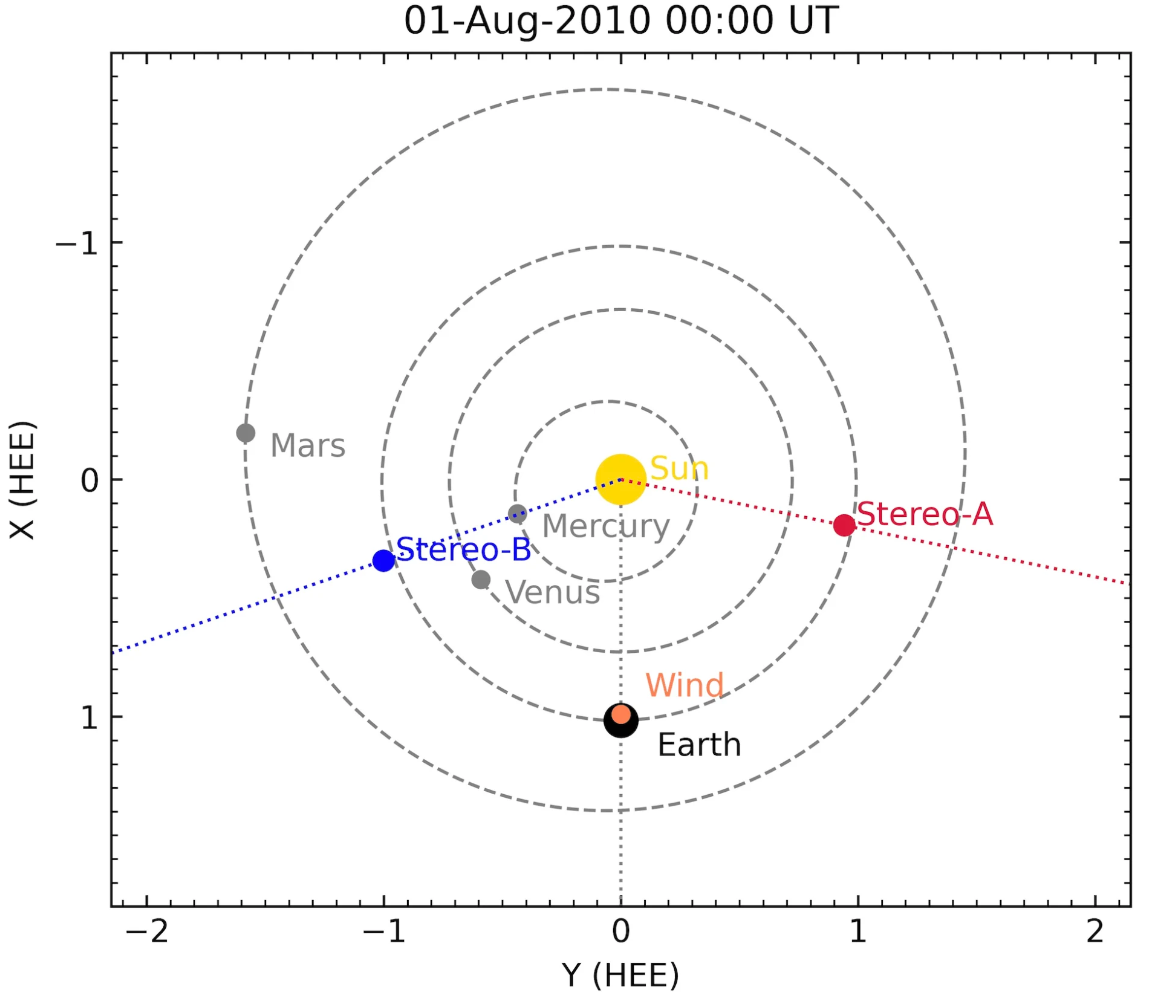}}
\caption{\small Heliocentric-Earth-Ecliptic (HEE) positions of the probes used in this research on the day of the prominence eruption under investigation. We include Wind at the Earth and note that ACE was at the same location. }
\label{fig:spacecraftlocations}
\end{figure}

The prominence eruption began at $\sim$05:00 UT on 2010 1 August \citep{Li2011}. This eruption is driven by a quiescent filament of particular interest because its eruption was captured at the Sun by instruments on three solar probes: the Atmospheric Imaging Assembly (AIA; \cite{Lemen2012}) on the Solar Dynamics Observatory (SDO), and the Solar TErrestrial RElations Observatory’s crafts A and B (STEREO-A and B; \citealt{Kaiser2005, Howard2008}), which all image the solar disk and low corona in various extreme ultraviolet (EUV) channels, (EUVI; \citealt{Wuelser2004}). At the time of the eruption the three spacecraft had significant angular separation (nearly 150$^{\circ}$) and were nearly equidistant to each other: STEREO-B was approximately 70.77$^{\circ}$ east of AIA and the Sun-Earth line, and STEREO-A was 78$^{\circ}$ west of AIA, as shown in Figure \,\ref{fig:spacecraftlocations}. These three vantage points constrain the off-limb evolution with independent estimates of the prominence's thermodynamic properties.

\begin{figure*}[h]
\includegraphics[width=0.33\linewidth]{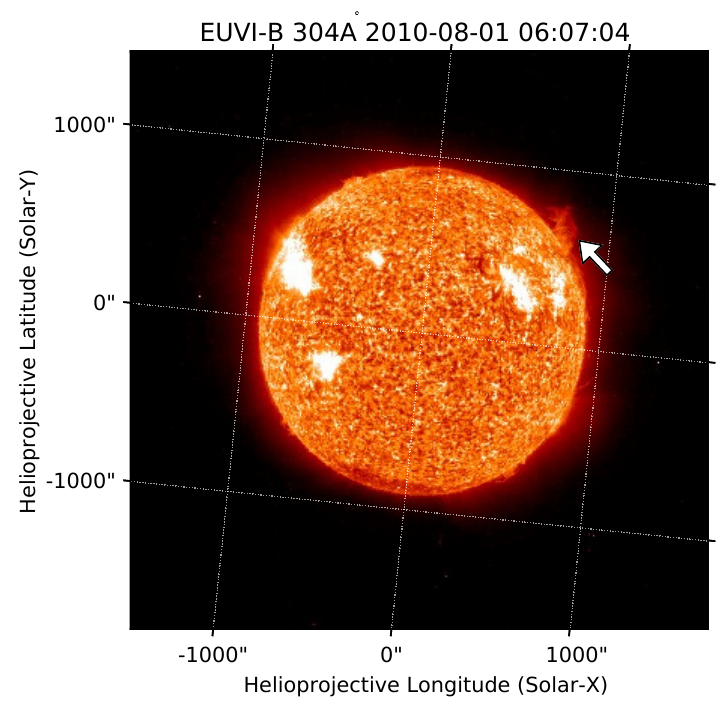}
\includegraphics[width=0.33\linewidth]{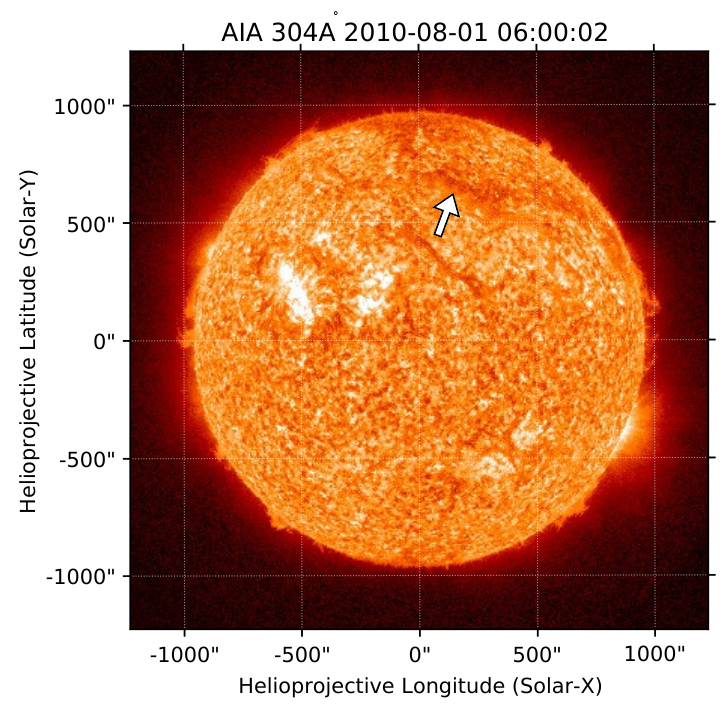}
\includegraphics[width=0.33\linewidth]{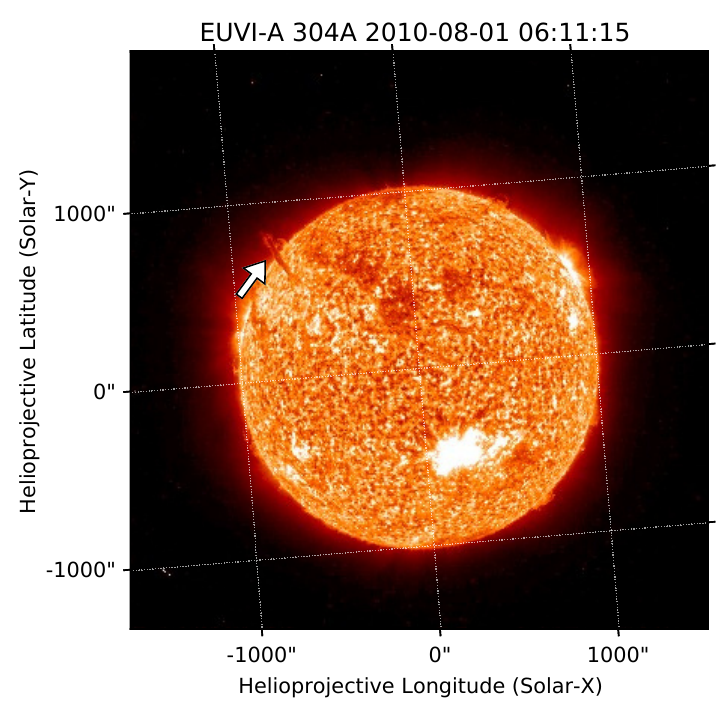}
\includegraphics[width=0.33\linewidth]{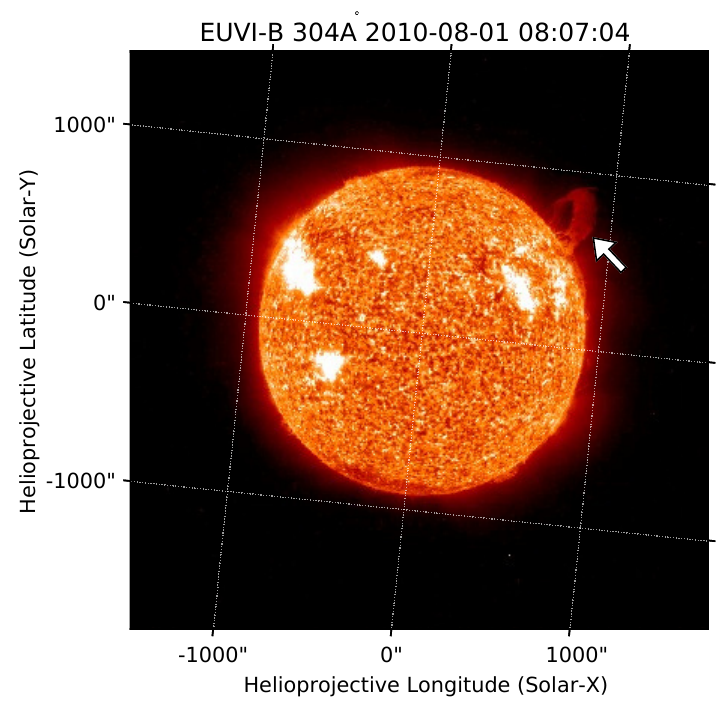}
\includegraphics[width=0.33\linewidth]{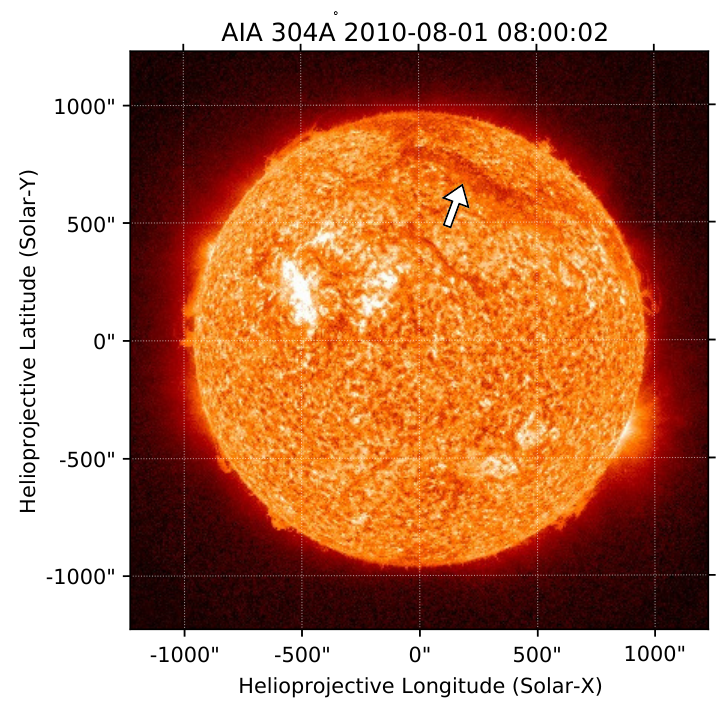}
\includegraphics[width=0.33\linewidth]{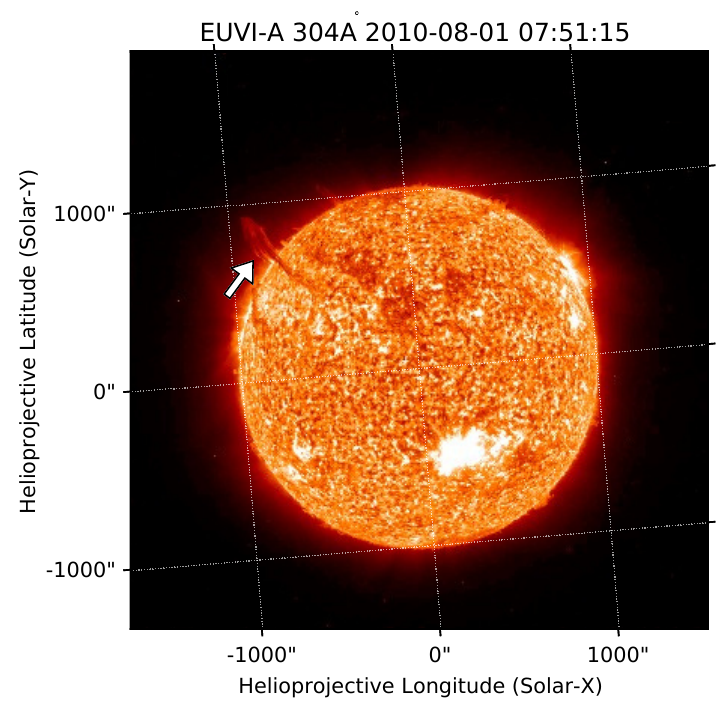}
\caption{\small From left to right, EUV images from the 304 \AA waveband of STEREO-B, the 304 \AA~ waveband of SDO/AIA, and the 304 \AA~ waveband of STEREO-A. The top row shows an early time in the eruption, while the bottom row shows a timestamp exactly two hours later.}
\label{fig:EUV_probe_images}
\end{figure*}

Figure \ref{fig:EUV_probe_images} shows the filament eruption in the three vantage points in the 304\AA~EUV channel. As noted, left to right is STEREO-B, SDO, STEREO-A, for two periods during the eruption. Although observations from all instruments are available, our analysis only includes SDO and STEREO-A when deriving temperature and density constraints. The reason is that STEREO-B observed the prominence dominantly in emission breaking the assumption that allows us to simplify the radiative transfer equation where the plasma emission was negligible, as discussed in Section\,\ref{subsec: constraints}. However, from the STEREO-A and SDO perspective, there are strong and continuous absorption off limb or on the disk, respectively, for this analysis. We discuss this implication in the discussion, Section \ref{sec:discussion}. However, we use both STEREO-A and B to derive an outflow velocity, where the prominence can be seen moving off-limb from both their perspectives. 

\subsection{In Situ Observations}
\label{sec:in-situ-observations}

\begin{figure*}[h!]
\centerline{{\includegraphics[width=0.9\textwidth]{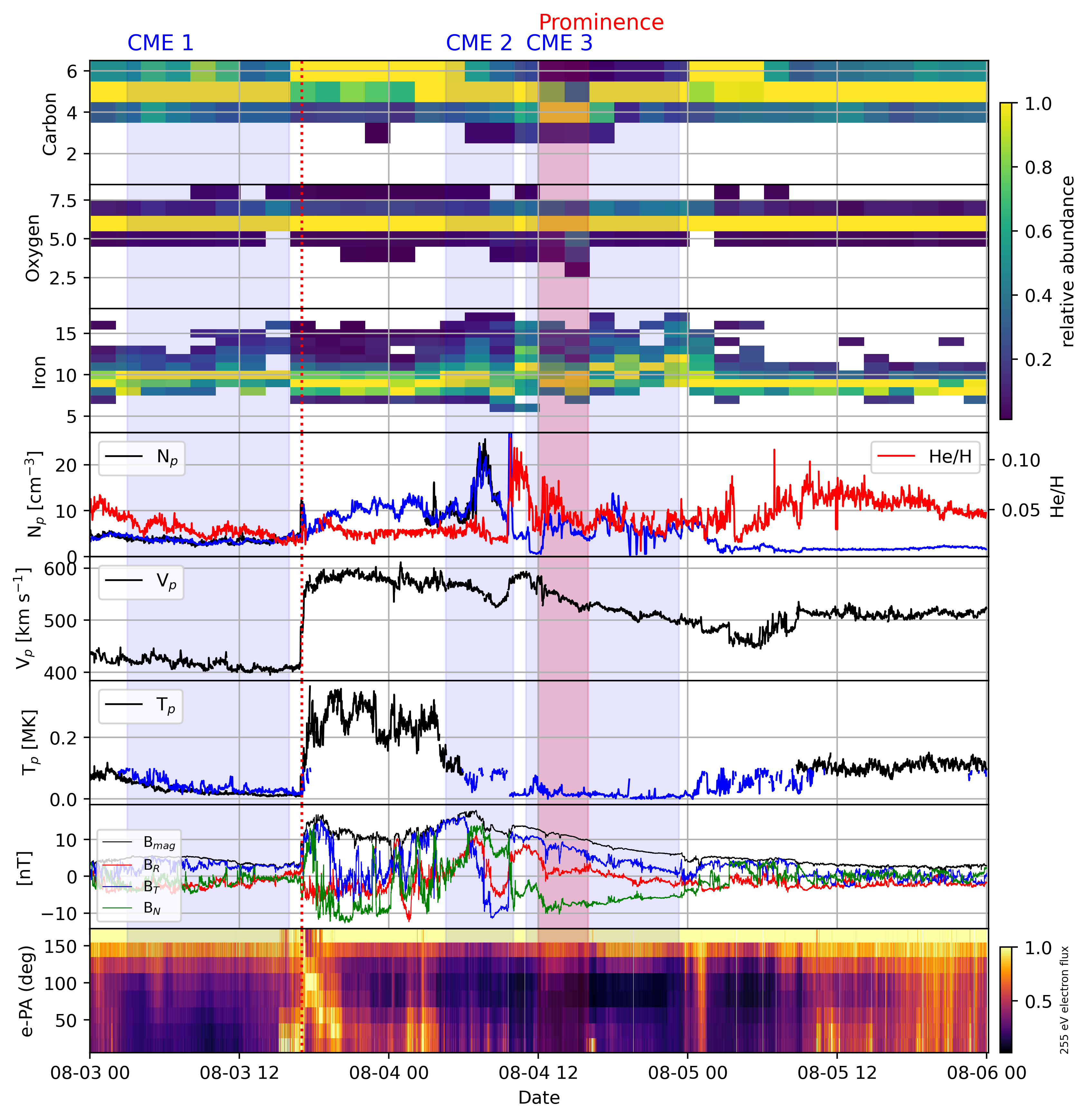}}}
\caption{\small From top to bottom, the figure displays the normalized charge state distributions of carbon, oxygen, and iron, proton density from ACE (black) and Wind (blue), as well as the He/H ratio (red), speed, proton temperature, magnetic field components and magnitude, and the last panel shows the column normalized electron strahl flux at 255 eV across pitch angle. The dotted red line marks the shock and start of the CME, the shaded blue regions are distinct CMEs identified in \cite{Liu2012, Mostl2012}, the shaded red region is the period of prominence material identified in \cite{Lepri2021}.}
\label{fig:Wind_data}
\end{figure*}

As noted by \cite{Lepri2021}, this eruption was also detected at the Earth-Sun L1 point by the Solar Wind Ion Composition Spectrometer (SWICS: \citealt{Gloeckler1992}) of the Advanced Composition Explorer (ACE: \citealt{Stone1998}) and by the Solar Wind Experiment (SWE; \citealt{Ogilvie1995}) Faraday Cup of the Wind spacecraft \citep{Wilson2021Wind}, providing a comprehensive set of constraints on the final state of the event far from the Sun. Figure\,\ref{fig:Wind_data} shows the passage of multiple CMEs from these spacecraft, as indicated by the blue shaded regions. The dashed red line shows the location of a shock following \textit{CME 1}, where there is a simultaneous enhancement of the density, speed, temperature and magnetic field magnitude observed near midday on August 3rd, also identified in \cite{Richardson2010} as well as in \cite{Liu2012, Mostl2012}. Critical to this study, the figure shows the carbon, oxygen, and iron relative abundances in the first three panels, followed by the density and He/H, speed, temperature, magnetic field, and electron pitch angle distributions where the density and temperature are a combination of the ACE (black) and Wind (blue) datasets. The shaded red region is the period associated with the prominence material embedded within the CME body, as identified in \cite{Lepri2021}. The prominence appears within a helical flux rope structure within \textit{CME 3}. Generally, there is a lower temperature for the duration of the prominence period as compared to the plasma detected immediately before and after 
\textit{CME 3}. There is also an enhancement in the helium abundance (red curve in the fourth panel) within the prominence period. We also note that the series of CMEs were also geoeffective, causing a geomagnetic storm as it passed the Earth, with a peak intensity/minimum Dst of -65 nT \citep{Wu2011}.

At the shock front, the charge state distribution of carbon shows an enhancement in C$^{6+}$ for $\sim 6$ hours, indicating strong heating happening immediately after the shock. The heating is also reflected in the broader charge state distribution of iron which shows an extension to higher ionization states. Strong heating of CME plasma is also observed directly in the high proton temperature compared to the pre-shock temperature. The prominence period shows significantly lower charge states compared to the surrounding CME where C$^{3+}$, C$^{4+}$, O$^{3+}$, O$^{4+}$ are measured. These lower charge states are not typically observed in the solar wind, only during CMEs. Congruent to the low charge states, the prominence period also contains highly ionized charge states, with a non-negligible contribution of to Fe$^{16+}$ for instance. These range of charge states provide some indication that at least part of the prominence undergoes significant heating. 

\section{Methodology}\label{sec:methods}

The work uses the methodology outlined in \cite{Landi2013} and implemented in \cite{Lee2017, Kocher2018}, which capitalizes on the absorption properties of the prominence to simultaneously estimate the temperature and density of prominence plasma. For our analysis, the derived quantities near the Sun serve as initial conditions to the temperature and density profiles as well as outflow speed, up to the ion freeze-in height. Beyond the low corona, we follow the prominence evolution by using profiles of temperature, density and speed derived from a previously constrained prominence eruption examined in \cite{Rivera2019a, Rivera2019b}. The study reconstructed the radial evolution of a 2005 CME using charge state distributions measured in the heliosphere. The work indicated the presence of a prominence, prominence–corona transition region, and hot surrounding outer CME component. The thermodynamic evolution of the distinct CME components from \cite{Rivera2019b} offers a set of thermodynamic profiles within different CME plasma structures that are utilized in this study to extend our profiles beyond our field of view. 

The derived initial conditions at the Sun were coupled with ion abundance measurements of the prominence material detected in situ during its passage at L1, provide constraints to the simulated carbon, oxygen, and iron ionization and recombination evolution. That is, the radial profiles of the prominence temperature, density, and speed were constrained by initial conditions at the eruption site and the profiles were iteratively adjusted to reconstruct the full carbon, oxygen, and iron charge states from its eventual passage near the Earth. 

The methodology consists of two main steps: First, in Section \ref{subsec: constraints}, we analyze EUV images from SDO and STEREO to derive the temperature, density, and velocity for the first few hours of the prominence eruption as it traveled outward from the Sun. Second, in Section \ref{subsec: model_plasma}, we use these constraints as initial conditions to extrapolate the plasma's temperature, density, and bulk speed profiles out from the Sun, and check they produce consistent bulk parameters at 1~au as compared to in situ data. These profiles are then directly used to drive a non-equilibrium ionization and recombination code that simulates the carbon, oxygen, and iron ion abundances out to their freeze-in altitudes. The simulated ions are compared in situ observations of the ion abundances to iteratively refine the plasma temperature, density, and velocity profiles and ultimately arrive at a self-consistent multi-thermal picture of the prominence which could not be inferred from currently available remote sensing alone.

\subsection{Deriving Temperature, Density, and Velocity Constraints From Remote EUV Observations} \label{subsec: constraints}

We examined several EUV broadband images of the eruption across the low corona to derive the column density and electron temperature of the absorbing plasma as it flows away from the eruption site. 

The prominence material in our present study is seen in absorption in several of the EUV images. Therefore, we can use a simplified version of the full radiative transfer equation (Eq. \ref{eq:full_rad}) to derive the properties of the plasma. 

\begin{equation}
F_{obs} = F_{inc} e^{-\tau} + F_{emitt} \\
\label{eq:full_rad}
\end{equation}

We assume the prominence is a non-EUV emitting slab of material and assume there is no scattering ($F_{emitt} = 0$), as in \citealt{Landi2013, Kocher2018}. In this case, we can simplify our equation to just the first term, that describes the attenuation of the background emission due to the prominence plasma. Therefore, our equation becomes:
\begin{equation}
F_{obs} = F_{inc} e^{-\tau}
\label{eq:f_obs_eq}
\end{equation}
Where $F_{obs}$ is the observed intensity, $F_{inc}$ is the background emission being attenuated, and $\tau$ is the optical depth defined by Equation \ref{eq:tau_keff}. $F_{obs}$ is taken to be the average pixel intensity measured within the prominence. $F_{inc}$ is an estimate of the average intensity of the background corona located near the prominence at each respective timestamp within the imager's FOV. The optical depth is defined as:

\begin{equation}
\tau = \int_{0}^{S} [\sum_{X} \sum_{m} n(X^{m+}) k_{X,m} \space] \space dl
\label{eq:tau_keff}
\end{equation}

The optical depth along the path 0 (at the prominence) to S (at the observer) is governed by all the constituents that can absorb along that path. $n(X^{m+})$ are the absorbers and $k_{X,m}$ are the absorption cross-sections of each constituent. In this case, $X$ is the element and $m+$ is the ionization stage of the ion. 
 
In the filament case, the main absorbing constituents are neutral hydrogen, neutral helium, and singly ionized helium (H, He, and He$^{+}$, or equivalently, H {\footnotesize I}, He {\footnotesize I} and He {\footnotesize II} respectively). Note that fully ionized hydrogen and helium do not participate as they do not contain bound electrons. Therefore, the optical depth simplifies to:

\begin{equation}
\tau = \int_{0}^{S} n_H k_{eff} \space dl
\label{eq:tau_simplified}
\end{equation}
where $k_{eff}$ is defined as:

\begin{equation}
\begin{aligned}
k_{eff} = f(\text{H {\footnotesize I}, T}) 
k_{\text{H {\footnotesize I}}} + A_{He} [f(\text{He {\footnotesize I}, T}) k_{\text{He {\footnotesize I}}} + \\ f(\text{He {\footnotesize II}, T}) k_{\text{He {\footnotesize II}}}]
\label{eq:k_eff}
\end{aligned}
\end{equation}

where $f(\text{H {\footnotesize I}, T}) = \text{H {\footnotesize I}/H}$ is the fraction of hydrogen which is neutral and a function of $T$, the electron temperature, $f(\text{He {\footnotesize I}, T}) = \text{He {\footnotesize I}/He}$ and $f(\text{He {\footnotesize II}, T}) = \text{He {\footnotesize II}/He}$ are the fraction of all helium species which are either neutral or singly ionized, and A$_{He}$ = He/H is the ratio of the total number density of hydrogen to that of helium.

Lastly, k$_{H{\footnotesize I}}$, k$_{He{\footnotesize I}}$, k$_{He{\footnotesize II}}$ are the absorption cross-sections of the corresponding ion. 

The Helium abundance of the plasma, which has photospheric abundance of 0.084 or 8.4\% \citep{Asplund2021}, has been well documented to spans a range of $1-5\%$ in the solar wind but can be as high as 30\% in some extreme cases, i.e. CMEs \citep{Borrini1982, Richardson2010, Alterman2019, Johnson2024}. For our event, the helium abundances range between 0.03 and 0.08, as shown in Figure \ref{fig:Wind_data}, so we use a constant value of 0.06 for the prominence.

The temperature-dependent fractional abundances, f(H {\footnotesize I}, T), f(He {\footnotesize I}, T), f(He {\footnotesize II}, T) from Eq. \ref{eq:k_eff}, can be computed assuming that our plasma is in ionization equilibrium. Ionization equilibrium assumes ionization and recombination processes of the system are in balance i.e. that the rate in which ions are produced and destroyed in the plasma are the equal so the overall change in the ion’s density is zero. This assumption is made only during the initial state of the eruption. In our quiescent case, collisional ionization equilibrium is valid in the filament given its high density in the corona. However, in the case where there is a flare, flare-related photoionization can become important which may lead to NEI effects quickly. The prominence eruption in the present study is not associated with a visible flare, therefore we assume photoionization is a minimal effect and ionization equilibrium holds for the initial outflow period.

We note that k$_{\text{eff}}$ is wavelength ($\lambda$) dependent. This dependence has step-like, discrete behavior due to the K-edges of the atoms and ions present \citep{Verner1996}. These K-edges are therefore a natural basis in which to break up the EUV-UV wavelength range ($0-1000$\AA) .

Equations \ref{eq:wavelengthranges1} - \ref{eq:wavelengthranges2} show the four main wavelength regions given the dominant absorbing species in our prominence.

Specifically, 912, 504, and 228\AA~ correspond to the K-edges associated with H{\footnotesize I} 13.6\,eV, He{\footnotesize I} 24.6\,eV, and He{\footnotesize II} 54.4178\,eV (where the eV values are the equivalent photon energy), as in \cite{Landi2013}: 

\begin{flalign}
    A: & 912 < \lambda < 1000: 
    k_{\text{eff}} = 0  \label{eq:wavelengthranges1} \\
    B: & 504 < \lambda < 912: k_{\text{eff}} = f(\text{H{\footnotesize I}, T}) \: k_{\text{H{\footnotesize I}}}\\
    C: & 228 < \lambda < 504: k_{\text{eff}} = f(\text{H{\footnotesize I}, T}) \: k_{\text{H{\footnotesize I}}} \nonumber \\ & + \text{A}_{\text{He}} \: f(\text{He{\footnotesize I}, T}) \: k_{\text{He I}} \\
    D: & 100 < \lambda < 228: k_{\text{eff}} = f(\text{H{\footnotesize I}, T}) \: k_{\text{H{\footnotesize I}}} \nonumber \\ & + \text{A}_{\text{He}} \: f(\text{He{\footnotesize I}, T}) \: k_{\text{He{\footnotesize I}}} + f(\text{He{\footnotesize II}, T}) \: k_{\text{He{\footnotesize II}}}
        \label{eq:wavelengthranges2}
\end{flalign}

All atomic properties are computed using atomic data from CHIANTI v11 \citep{delzanna2021} as well as using version 0.2.3 of the fiasco open source software package \citep{Barnes2024}.

To solve for the column density $N_H$, or $L(T)$ (as referred to in \citealt{Landi2013}), of the material, we can rearrange Equation \ref{eq:f_obs_eq} as:

\begin{equation}
N_H = L(T) = \frac{ln(F_{obs}/F_{inc})}{-k_{\text{eff}}(A_{\text{He}}, T)} \label{eq:simplified_f_obs_eq}
\end{equation}

The column density, $N_H$, is the integral of the total number density of hydrogen, $n_H$ along a given line of sight. 

\begin{figure}[h!]
\centerline{\includegraphics[width=0.4\textwidth]{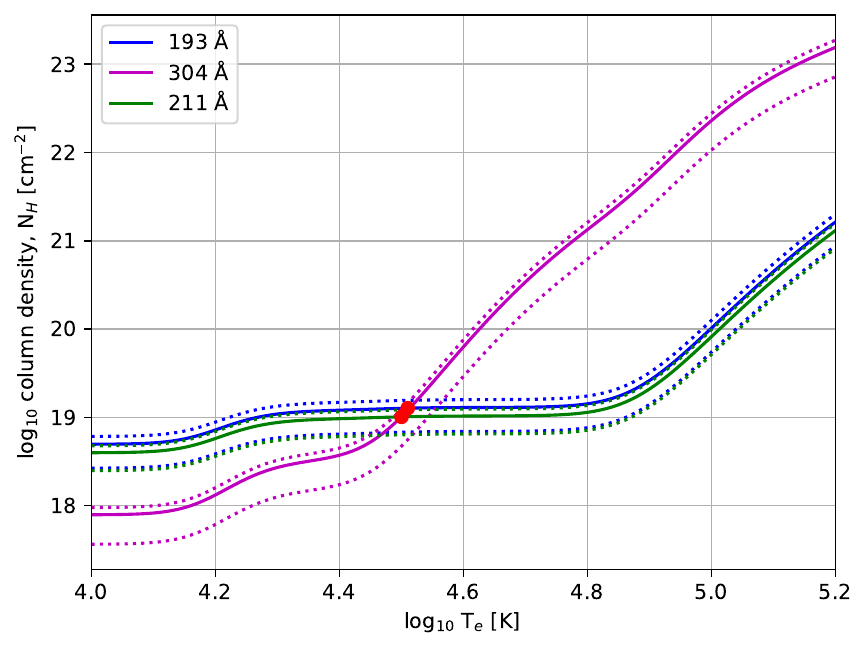}}
\caption{\small Illustration of the log of column density, L(T) or n$_{H}$, with increasing electron temperature computed using Equation \ref{eq:simplified_f_obs_eq} using SDO/AIA EUV emission during a snapshot of the prominence evolution. The intersection of the column density curves of different wavelength regions is used to determine the temperature and column density of the observed plasma.}
\label{fig:ltplot}
\end{figure}

Plotting the column density as a function of the electron temperature, given by Equation \ref{eq:simplified_f_obs_eq} within at least two wavelength ranges in Equations \ref{eq:wavelengthranges1} - \ref{eq:wavelengthranges2}, jointly solves for the column density and the electron temperature. Figure \ref{fig:ltplot} illustrates this procedure, showing the log column density and electron temperature inferred as the intersection point of three L(T) curves derived from the absorption in three wavelength AIA channels which probe $k_{eff}$ groups C (193, 211) and D (304). To account for the variability in the measured column density values that result from applying this methodology to several wavelengths in the same $k_{eff}$ group, we use the mean value of the different intersection points. 

\begin{figure*}
{\includegraphics[width=0.95\textwidth]{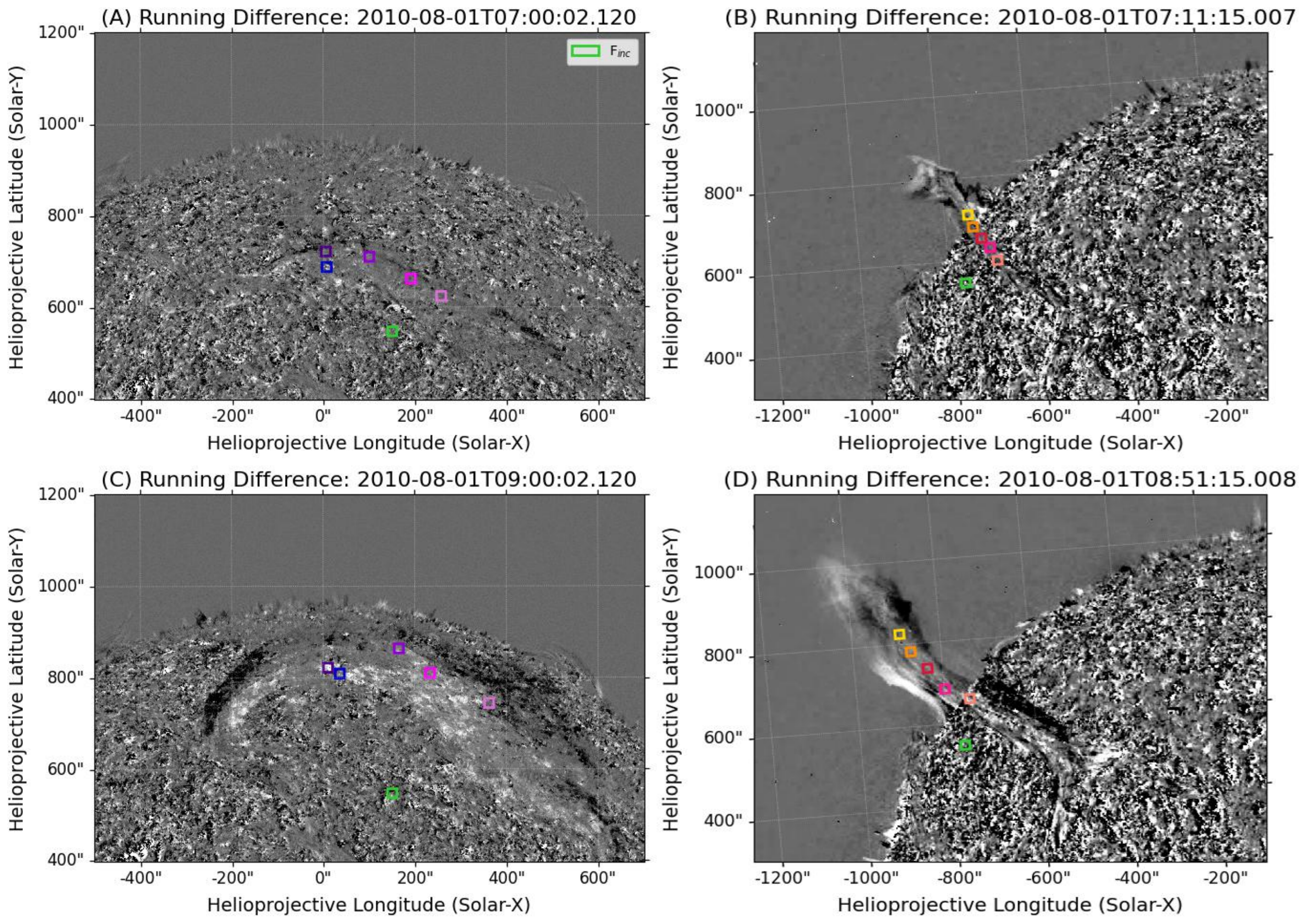}}

\caption{\small Two snapshots of running difference images, displaying the tracks used for AIA (A and C) and STEREO-A (B and D) at a time early on in the prominence's eruption (A and B), and again after two hours of evolution (C and D). The green 20'' box in each image is a example of a box of non-prominence material we used to calculate F$_{inc}$ values.}
\label{fig:AIA_and_STEREOA_tracks}
\end{figure*}

\begin{figure*}
{\includegraphics[width=0.95\textwidth]{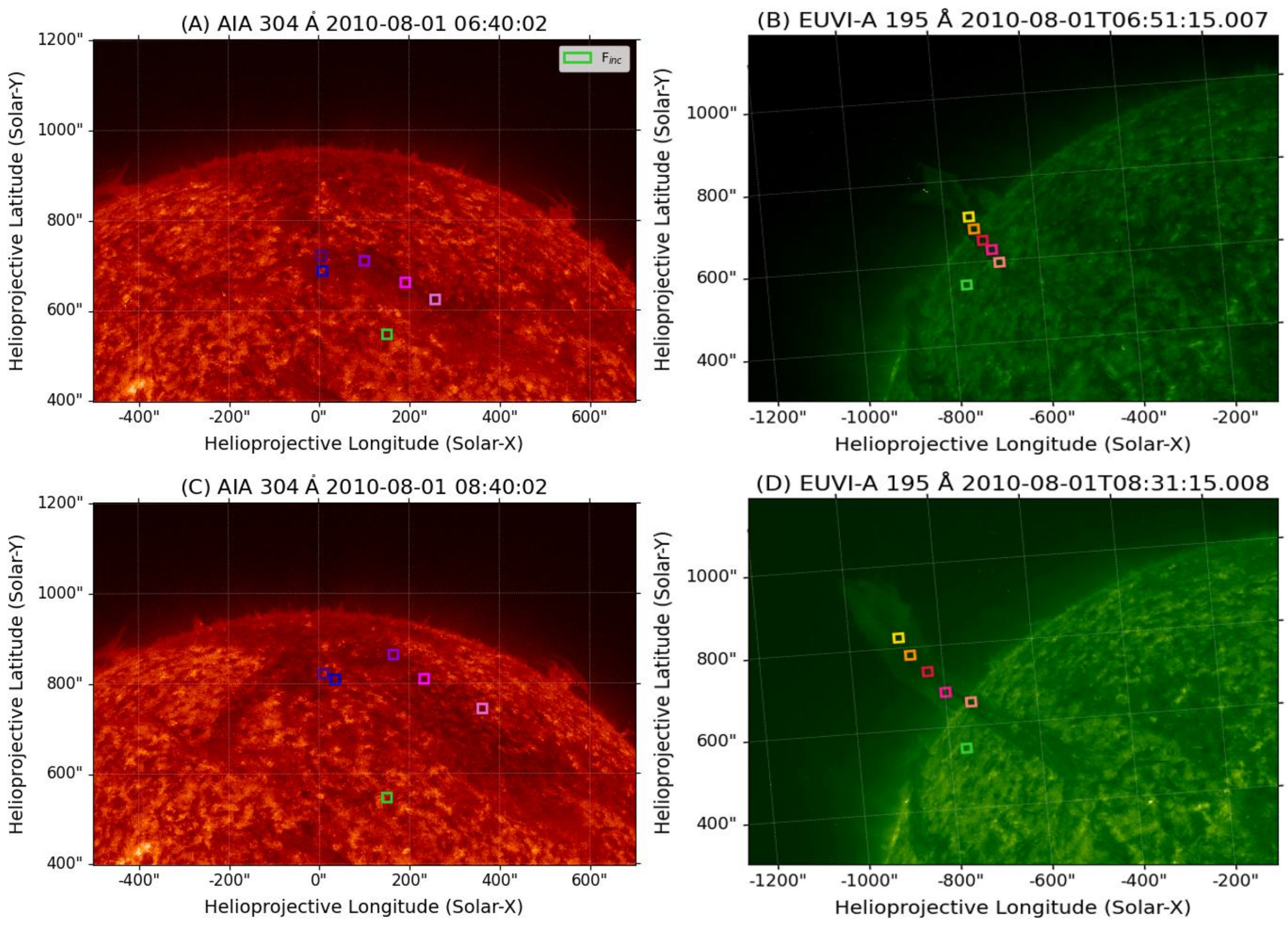}}

\caption{\small The same snapshots as Figure \ref{fig:AIA_and_STEREOA_tracks} are show in EUV.}
\label{fig:AIA_and_STEREOA_tracks_in_EUV}
\end{figure*}

To derive the density and temperature within the prominence as it erupts, we follow the changing F$_{obs}$ and F$_{inc}$ in time with boxes that follow the prominence evolution, as shown in Figures \ref{fig:AIA_and_STEREOA_tracks} and \ref{fig:AIA_and_STEREOA_tracks_in_EUV}.
For each box we measure F$_{obs}$ as the mean pixel intensity in the box and F$_{inc}$ from a box containing only quiet sun material (green boxes in the figures). We use these measurements to generate individual plots of Figure \ref{fig:ltplot} and identify an intersection point within each box in time. 
Figure \ref{fig:AIA_and_STEREOA_tracks} shows difference images of sequential observations where black (white) colors indicate pixel intensities decreasing (increasing) relative to the prior images. In the present context of a structure in absorption, this means black (white) colors shown the leading (trailing) edge of the prominence as it moves. 

As shown in both figures, there are $20''\times20''$ boxes along the prominence structure that we sequentially shift as the prominence moves away from the Sun. The starting points (boxes in panels A and B in both figures) of each track were selected based on the pre-eruptive structure while on the disk. Difference images were used to highlight the movement of prominence material and to identify the new location of each box of material in successive frames. In this way, whole tracks were followed from one image to the next by tracking the movement of these white patches over successive difference images. The loop of the material maintained a fairly self-similar shape within the timeframe we evaluated, and difference images (and therefore tracks of measurement) reflected this consistent structure.

For each box, we therefore obtain a value of the column density and the electron temperature as a function of time. We convert column density values ($N_H$, cm$^{-2}$) to number density ($n_H$, cm$^{-3}$) by dividing the value by the approximated width of the prominence along the LOS, $1.453\times10^{9}$ cm, which amounts to assuming that $k_{eff}$ is constant over the portion of the prominence the LOS intersects. We estimate the size of the prominence as the width observed on the plane of the sky in the AIA or EUVI images during the different stages of evolution. 

With estimates of the plasma density and electron temperature in hand, the thermodynamic characterization of the prominence was completed by estimating its outflow speed. To do this, we used box positions in successive images in the STEREO-A 194\AA\ waveband (Figure \ref{fig:AIA_and_STEREOA_tracks_in_EUV} B, D). Velocity values were calculated by assuming the prominence traveled along a path of constant longitude and latitude (i.e. radially). We first computed the apparent speed of each part of the filament from their image coordinates yielding values in arcseconds per minute. Informed by the stereoscopic observations we then approximated the prominence path to be radial and anchored at 36.33 degrees from the plane of the sky. Dividing by the cosine of this angle and approximating 1 arcsecond as 727\,km, we obtain outflow speeds of order 10s of km/s which are in line with the velocity profile of prominence material studied in \cite{Kocher2018}, and with the velocities measured by \citealt[][see their Figure 7]{Li2011}. We track the box positions with this same general radial assumption to associate an altitude with each box position. 

As will be presented in Section \ref{sec:results}, at this stage we are able to fully characterized the prominence's thermodynamic evolution in the low corona with approximate altitude, plasma density, outflow speed and electron temperature tabulated for various sampled points on the prominence as a function of time.

\subsection{Ionization and Recombination Modeling} \label{subsec: model_plasma}

With the methodology for determining the thermodynamic properties of the prominence eruption low in the corona established, we now explain how this information is used to predict how the plasma should continue to evolve out to each ion's freeze-in distance. To do this, the temperature, density, and velocity conditions in the low corona are used as the initial conditions to adjust radial thermodynamic profiles derived in \cite{Rivera2019a}. These profiles in turn are used to simulate the non-equilibrium ionization (NEI) and recombination of the charge state history of carbon, oxygen, and iron in the plasma.

Given radial profiles of speed ($U$), density ($n_e$) and temperature ($T_e$), the NEI model follows \cite{Shen2015}, with ion abundance fractions obeying:

\begin{equation}
    \frac{df_k^{i}}{dt} + U \frac{d}{ds} f^i_k = n_e[C_k^{i-1} f^{i-1}_k - (C^i_k +R^{i-1}_k)f^i_k + R^i_kf^{i+1}_k]
\label{eq:fractional_abundance_eq}
\end{equation}

where $C_k$ and $R_{k}$ are rate coefficients for the ionization or recombination of the k-th charge state of an ion, and $f_k$ refers to the fractional abundance of that charge state, and are all functions of the local $T_e$. Using the implicit method equation (Equation \ref{eq:implicit_method_frac_abundance}), outlined in Appendix A of \citep{Barnes2024}, we solve Equation \ref{eq:fractional_abundance_eq} for the case where the plasma is out of ionization equilibrium.

\begin{equation}
\mathbf{F}_{l+1} = \left(\mathbb{I} - \frac{\delta}{2}\mathbf{A}_{l+1}\right)^{-1}\left(\mathbb{I} + \frac{\delta}{2}\mathbf{A}_l\right)\mathbf{F}_l \label{eq:implicit_method_frac_abundance}
\end{equation}

where $\mathbf{F}_l$ is the vector of population fractions of all states at time $t_l$, $\mathbf{A}_l$ is the matrix of ionization and recombination rates with dimension $Z+1\times Z+1$ at time $t_l$, and $\delta$ is the time step.

NEI effects are particularly important to capture charge state evolution in processes with rapid changes in the plasma conditions, as is the case for flare and CME evolution where the temperature, density, and bulk speed of the plasma vary quite suddenly. NEI modeling has been implemented extensively in the past to 1-D models of the solar wind \citep{Landi2012, Gilly2020, Rivera2020} and CMEs \citep{Rakowski2007, Gruesbeck2012, Rivera2019a, Laming2023} and coupled to 3D MHD solutions of the solar wind and CMEs \citep{Oran2015, Lionello2019, Szente2022, Lionello2023, Rivera2023, Riley2025, Wraback2025}. 

Assuming an initial state of ionization equilibrium, we compute the ongoing and cumulative ionization and recombination processes in the plasma and produce the resulting frozen-in charge state distributions to compare with the in situ measured distributions which were presented in the top three panels of Figure \ref{fig:Wind_data}. The agreement between the two distributions is evaluated with a cost function defined as:

\begin{equation}
{f(x)}^2 = \sum_{i = 1}^{n} {\frac{(O_i - M_i)^2}{M_i}}
\label{eq:cost_function}
\end{equation}

Here $n$ is the number of ions in the histogram that the function has to evaluate over, $O_i$ is the observed in situ relative abundance of each ion, and $M_i$ is the relative abundance of each ion from the mix of model plasmas. The most constraining ion we evaluate is iron with $n=11$ populated charge states. 

Informed by the bi-modal distribution of iron charge states observed at 1~au, we quickly determine that a single temperature population is unable to reproduce the frozen-in population. Therefore, we introduce a two temperature population into our procedure. That is, we take our system to be composed of a majority cool and minority hot component with the former following the temperature profile of cool prominence material \citep[profile PC1 in ][]{Rivera2019a} and the latter following that of hot surrounding material \citep[profile PC4 in ][]{Rivera2019a}. We fix the relative density of the hot and cool component with a 70:30 ratio (the approximate ratio of in situ charge state abundance above Fe XIII to below) and then vary the peak temperature of the PC1 and PC4 temperature profiles (keeping the initial condition of the cool component within the range of values determined by our remote observational analysis). The respective density and velocity profiles fixed and constrained by the remote observations.  

We vary these two peak temperatures systematically and determine the combination which yields the minimum of the cost function introduced above. From this parameter search we produce a self-consistent thermodynamic model for the prominence which most closely resembles the measured plasma composition, as presented and further discussed next in Section \ref{sec:results}. 

\section{Results}
\label{sec:results}

\subsection{Derived Temperature, Density, and Velocity Constraints From Remote Observations}

\begin{figure}[h!]
{\includegraphics[width=0.47\textwidth]{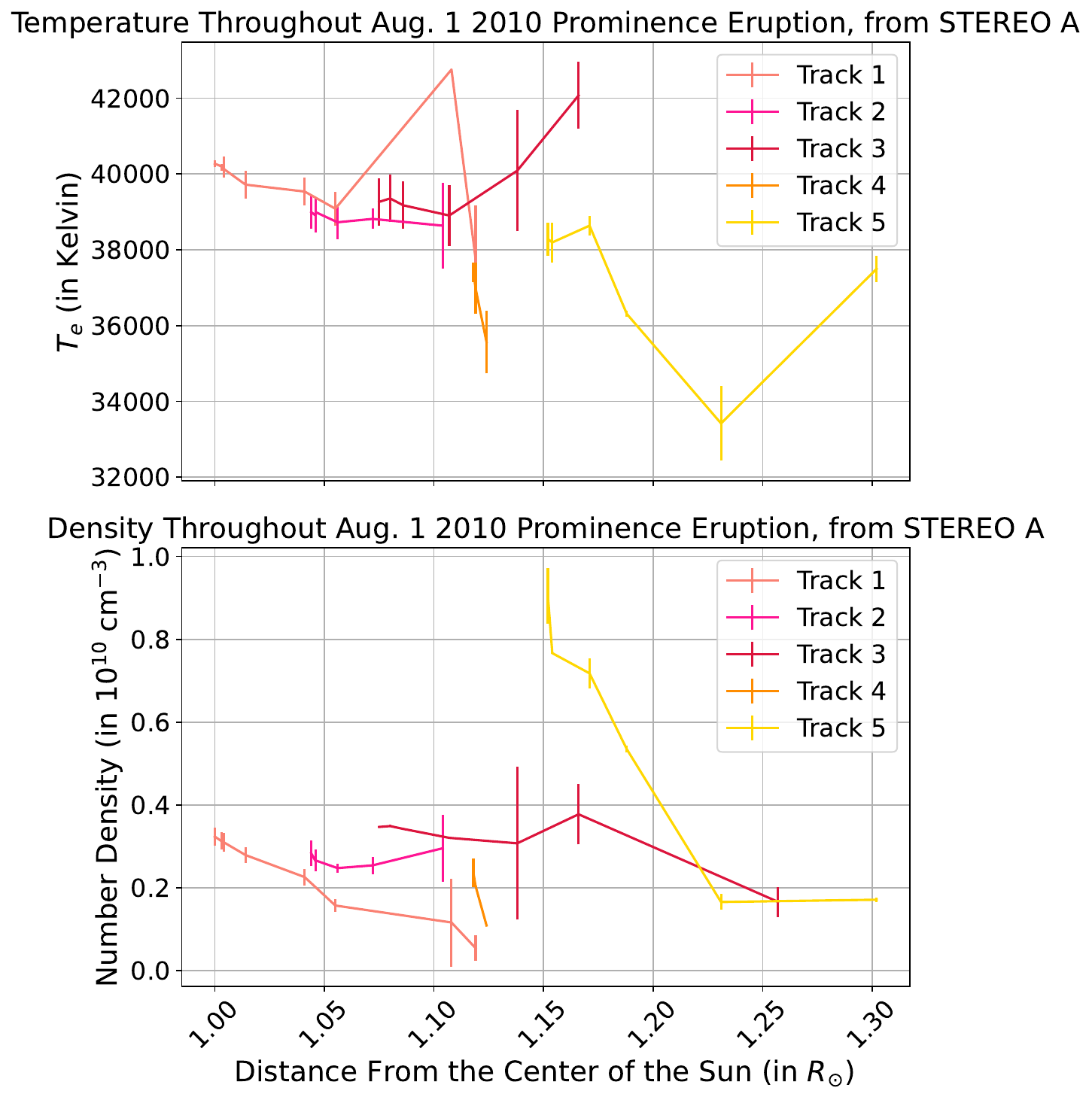}}
\caption{\small Plots displaying the temperature (top) and density (bottom) recorded across various tracks from STEREO-A as a function of distance from the center of the Sun. The colors of each box in Figure \ref{fig:AIA_and_STEREOA_tracks_in_EUV} correspond to the colors of the curves.}
\label{fig:functionofdistance}
\end{figure}

Figure \ref{fig:functionofdistance} shows the computed temperature (top panel) and column density (bottom panel) with distance from the Sun of the individual tracks across the prominence from the STEREO-A perspective, that are color coded according to the corresponding box colors in Figure \ref{fig:AIA_and_STEREOA_tracks}. The distance is harder to compute for the AIA tracks given the position of the observer so is omitted here. From these plots we can see that the density of the prominence material observed by STEREO-A only slightly declines as it moves away from the Sun, while the material's temperature does not follow a clear decline, but rather stays fairly consistent while near the Sun for most tracks. Track 5 which is from the apex of the loop is the exception to this, being further from the Sun and somewhat cooler, however it is unclear if this is due to inhomogeneity in the prominence, issues with the assumption of absorption off the disk, or of the other portions of the prominence would later experience a similar cooling. 

\begin{figure}[h!]
{\includegraphics[width=0.47\textwidth]{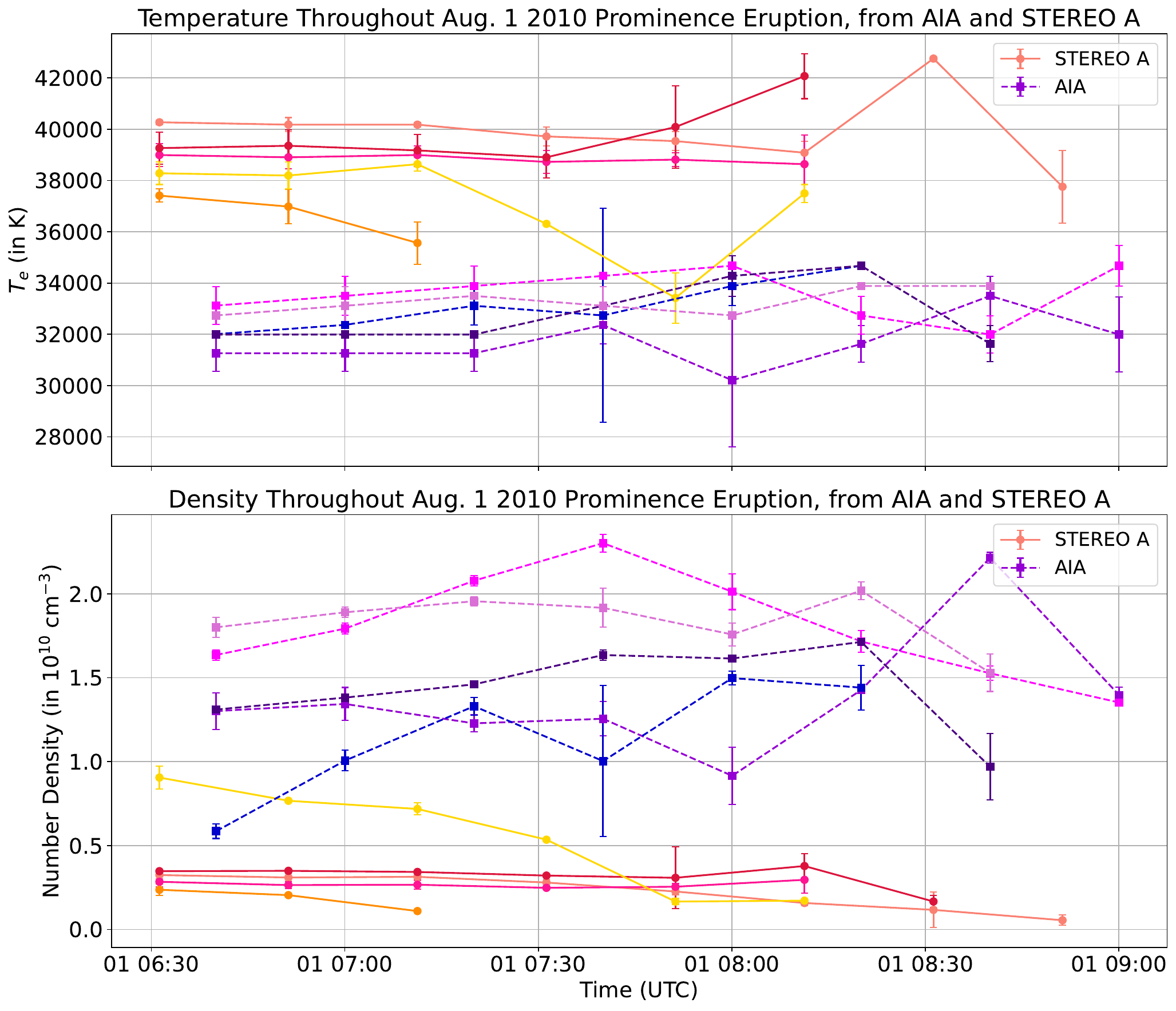}}
\caption{\small Plots of temperature (top) and density (bottom) as a function of time for both AIA and STEREO-A. STEREO-A's tracks noted consistently higher temperature and lower density than AIA's tracks, though the data from both sets of tracks were close and in line with the values we expect of a quiescent prominence eruption. The slight discrepancy in the two data sets is likely due to the head-on viewing angle of AIA, which makes discerning exact density values more difficult.}
\label{fig:functionoftime}
\end{figure}

Figure \ref{fig:functionoftime} shows the combined temperature (top) and density (bottom) of the prominence in time from both STEREO-A/EUVI and SDO/AIA. Evaluation of the plasma's thermal evolution reveals that temperatures remained relatively constant for the entire span of time the prominence material was within the field of view of the two imagers. It is worth noting that AIA appears to observe sudden temperature drop off towards the end of its observations of the eruption, while these temperature dips are still present, but less extreme, in the temperatures noted from STEREO-A.

The density profile of each track derived from both EUV imagers is slowly decreasing according to STEREO A (as seen also in the distance plot) but remains relatively constant according to AIA. The observations also differ from each other by almost an order of magnitude. In light of this discrepancy, it is worth nothing that AIA's calculated density values may have been effected by its head on viewing angle. This angle allows for prominence material ejected at later times to show up in the same box of prominence material we hope to track, causing a "pile up" of material that could impact measured values of column density. Still, all density values remain within a reasonable range and, assuming subsequent spherical expansion, are consistent with the extrapolated density values measured by the in situ observations of the same structure, as will be discussed in Section \ref{subsec: constraints}.

It is worth noting that since our methodology relies on the instrument's on-disk viewing angle (since we require that $F_{obs} >> F_{inc}$ for Equation \ref{eq:simplified_f_obs_eq} to hold), we could not continue recording temperature and density values once the prominence became significantly off-disk in the frames on AIA and STEREO-A, limiting our temperature and density constraints to the first approximately two hours of the prominence eruption which corresponds to the apex of the prominence reaching 1.25$R_\odot$.

\subsection{Plasma Ionization and Recombination Modeling}

As mentioned earlier, preliminary trials of the plasma ionization and recombination evolution used a single temperature plasma following the PC1 thermodynamic profiles of \citet[][]{Rivera2019a} with initial values fitted to the plasma's derived near-Sun temperature, density, and velocity determined from remote observations and evaluated this against the ACE/SWICS in situ charge state constraints for iron, carbon and oxygen (first three panels of Figure \ref{fig:Wind_data}.
Similar to \cite{Gruesbeck2012, Rivera2019a}, we were not able to capture the full range of carbon, oxygen, and iron charge states by modeling the properties of a single structure but required multi-thermal components, a cool and a hot plasma parcel, consistent with prior remote-sensing inferences of multi-thermal plasmas in the vicinity of solar prominences \citep{Habbal2010_hotshrouds}. Combining these initial conditions with non-ionization equilibrium modeling including heating of both profiles in the low corona, we are able to better account for the full range of observed charge states. 

\begin{figure}[h!]
\centerline{\includegraphics[width=0.5\textwidth]{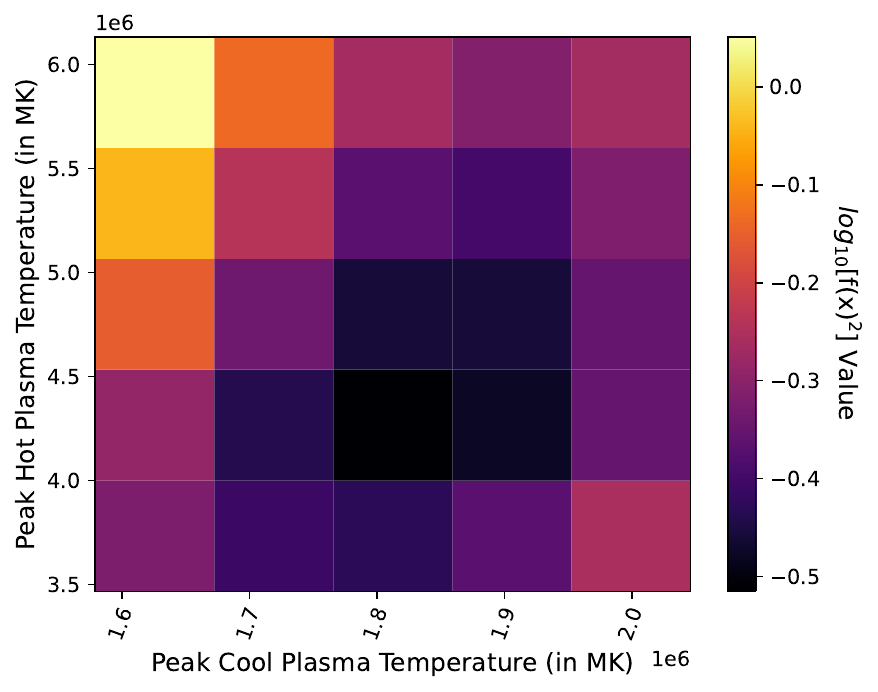}}
\caption{\small A color map of the cost function values of ion abundance model histograms with various cool plasma and hot plasma peak temperatures. The best fit cost function value of 0.305 corresponds to a dark purple color on the color map, while the poorest fit cost function  value of 1.123 corresponds to a light yellow color.}
\label{fig:2Dcolormap}
\end{figure}

The optimization process is illustated with the 2D histogram in Figure \ref{fig:2Dcolormap} showing the cost function as a function of the peak temperature of the hot and cool populations. The figure shows the cost function parameter space created by varying the temperature profiles of both the hot and cool plasma, keeping their density and velocity profiles aligned with their bookend constraints and a ratio of 70:30 by abundance. As the figure shows, the lowest cost function value is 0.305 obtained for this 70:30 ratio and peak temperatures of 1.8 and 4.3 MK respectively. Other abundance ratios were also tested and resulted in a larger minimum cost function value.

\begin{figure*}[h!]
\centerline{\includegraphics[width=0.5\textwidth]{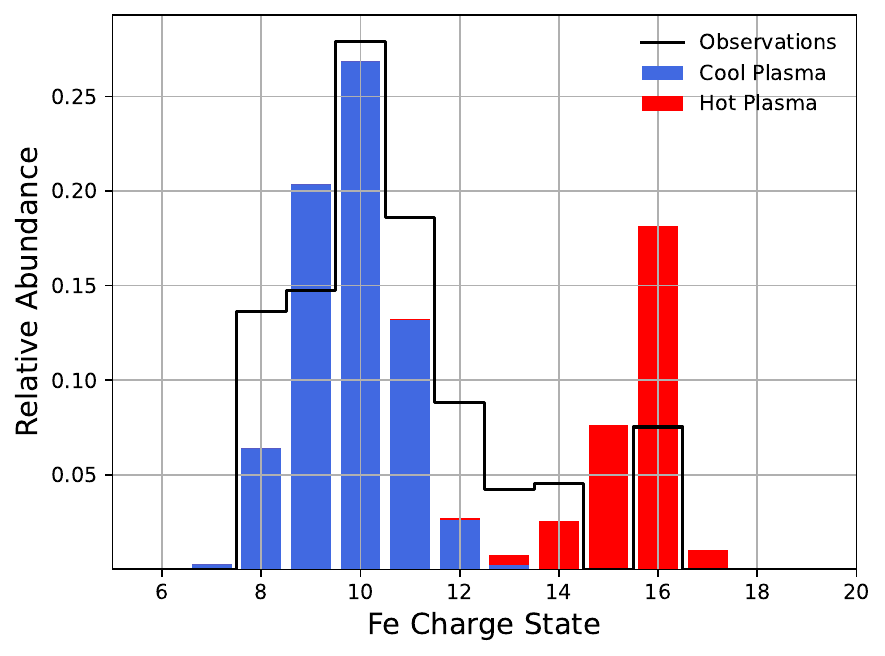}}
\centerline{\includegraphics[width=0.5\textwidth]{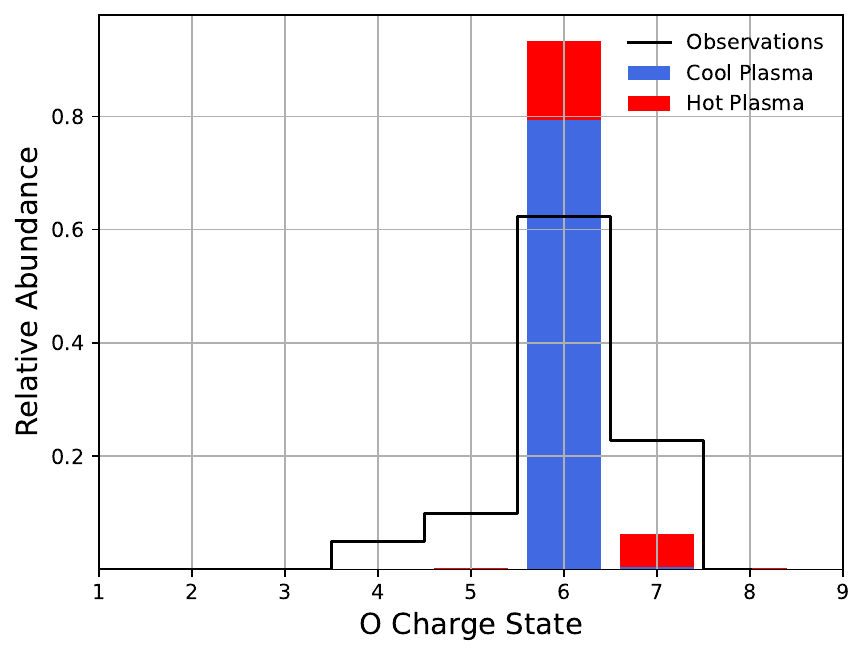}}
\centerline{\includegraphics[width=0.5\textwidth]{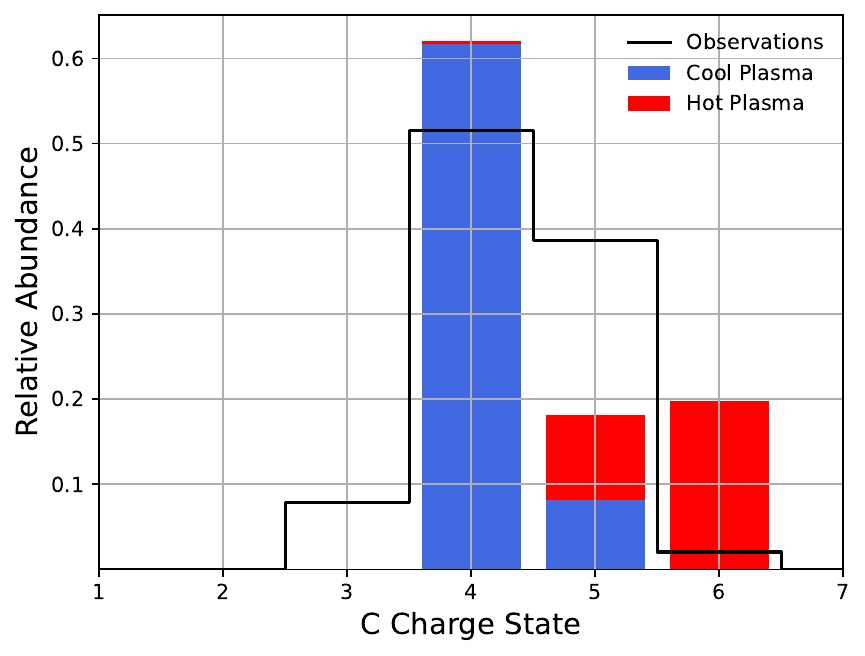}}
\caption{\small The iron fractional abundance histogram (top) for the combination of hot and cool plasma with the lowest cost function value, a value of 0.235. The oxygen (middle) and carbon (bottom) fractional abundance histograms corresponding to the same conditions as the iron histogram are also displayed. The double-peak in iron's fractional abundance observed by ACE was replicated by the model plasmas, along with the single peaks in oxygen and carbon abundance observed by ACE.}
\label{fig:best_fit_histogram}
\end{figure*}

The final distributions of fractional abundances for the combined hot/cool plasma is shown in Figure \ref{fig:best_fit_histogram}. In Figure \ref{fig:best_fit_histogram}, we can see how the hot and cool model plasmas contribute to the model prominence material's overall fractional abundances and clearly see that if a one component cool plasma is used, then none of the high charge states are reproduced. With the two component plasma, the contributions reproduce the double-peaked fractional abundance of iron observed, as well as matching the single peaks of both oxygen and carbon's observed fractional abundances well. However, apart from the iron distribution, both the simulated carbon and oxygen do not capture the lowest charge states observed e.g. C$^{3+}$, O$^{4+}$, O$^{5+}$. 

\begin{figure*}[h!]
\centerline{{\includegraphics[width=0.9\textwidth]{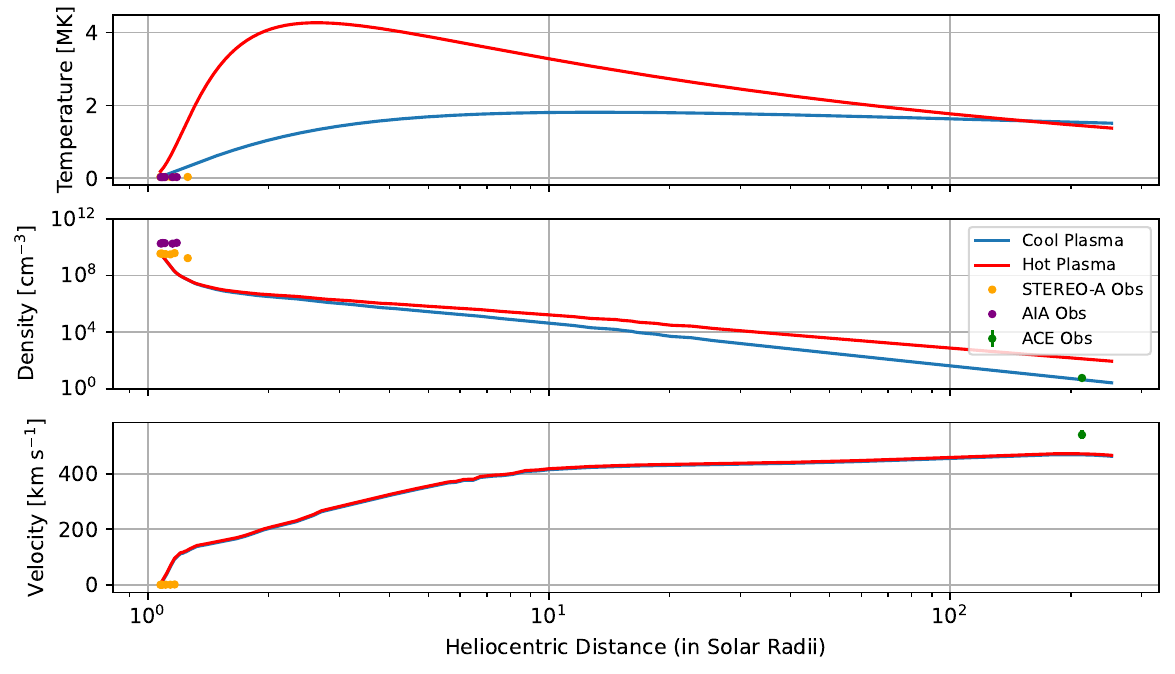}}}
\caption{\small The temperature, density, and velocity profiles from the Sun of the cool plasma (in blue) and the hot plasma (in red). Orange (STEREO-A) and purple (AIA) data points are derived prominence properties at the Sun. The green data points are measurements of the prominence recorded by ACE.}
\label{fig:plasma_profiles}
\end{figure*}

Figure \ref{fig:plasma_profiles} summarizes the final radial profiles used, labeled as 'Hot' and 'Cool' plasma, which include the derived coronal properties in orange and purple at the Sun (from STEREO and AIA where available, respectively) and in situ properties from ACE at L1. As mentioned previously, the 'Hot' component was taken from \cite{Rivera2019a}, their PC 4 profile and the cool profile from their PC 1 profile. As can be observed, the initial and final conditions of the cool plasma are quite close to the observational constraints at both ends of the profiles, suggesting we have derived a self consistent picture of the thermodynamic evolution of this plasma.  

\begin{figure}
\includegraphics[width=0.5\textwidth]{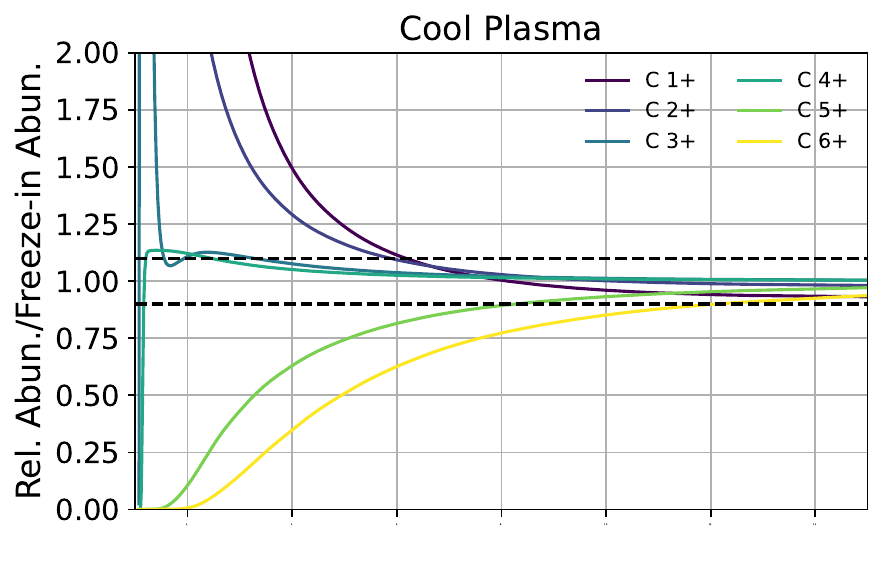}
\includegraphics[width=0.5\textwidth]{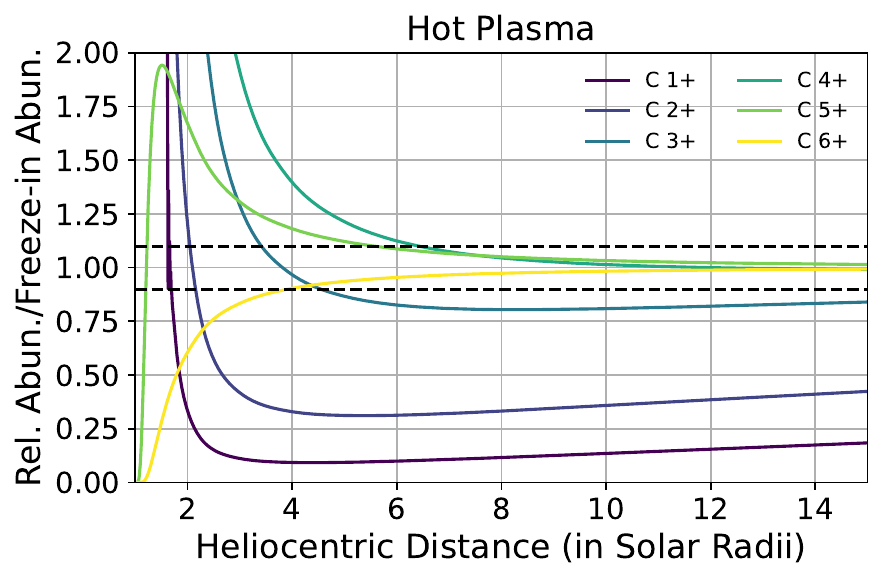}
\caption{\small Freeze-in evolution of carbon charge states for the cool plasma (top) and the hot plasma (bottom), from Figure \ref{fig:plasma_profiles}. Dashed horizontal lines are at 0.9 and 1.1, where the curve is within 10\% of 1.}
\label{fig:freeze-in-plot_c}
\end{figure}

\begin{figure}
\includegraphics[width=0.5\textwidth]{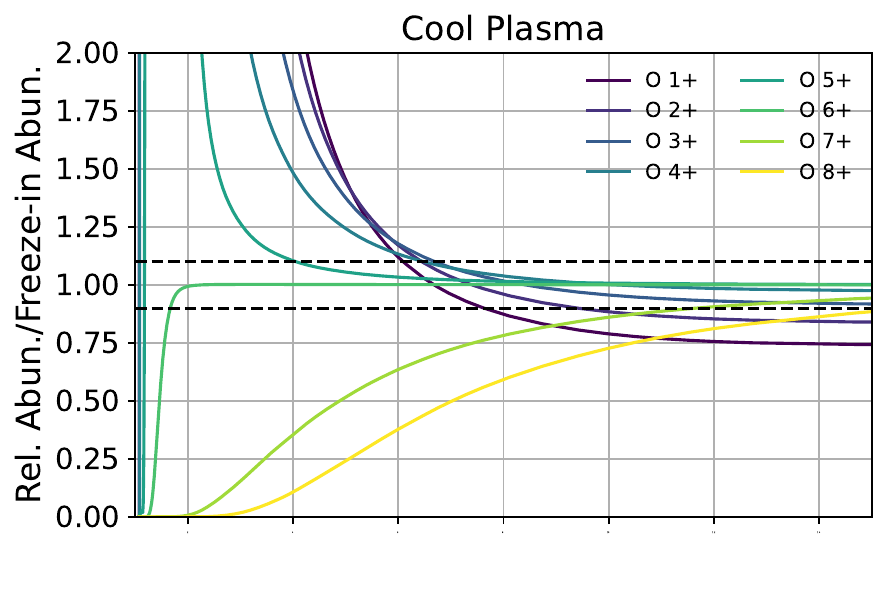}
\includegraphics[width=0.5\textwidth]{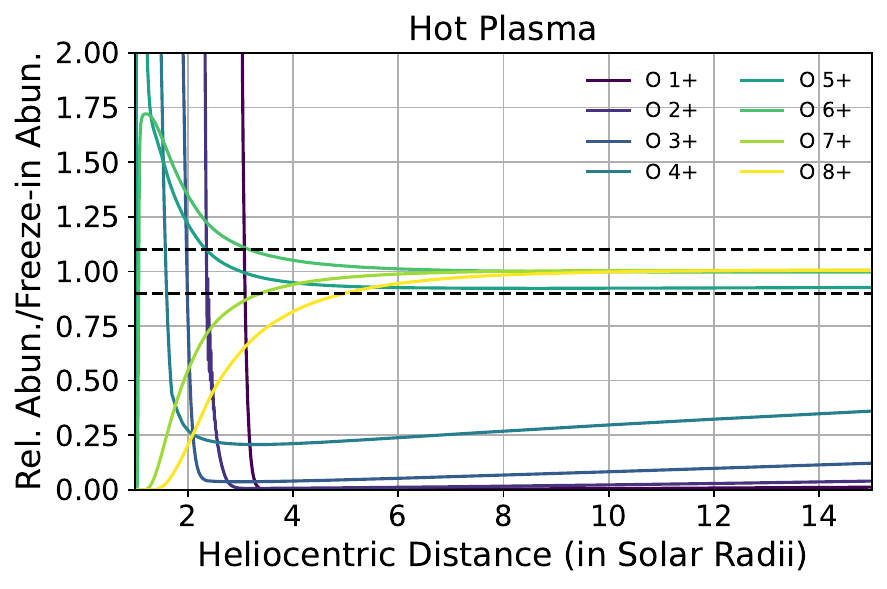}
\caption{\small Same as \ref{fig:freeze-in-plot_c}.}
\label{fig:freeze-in-plot_o}
\end{figure}

\begin{figure}
\includegraphics[width=0.5\textwidth]{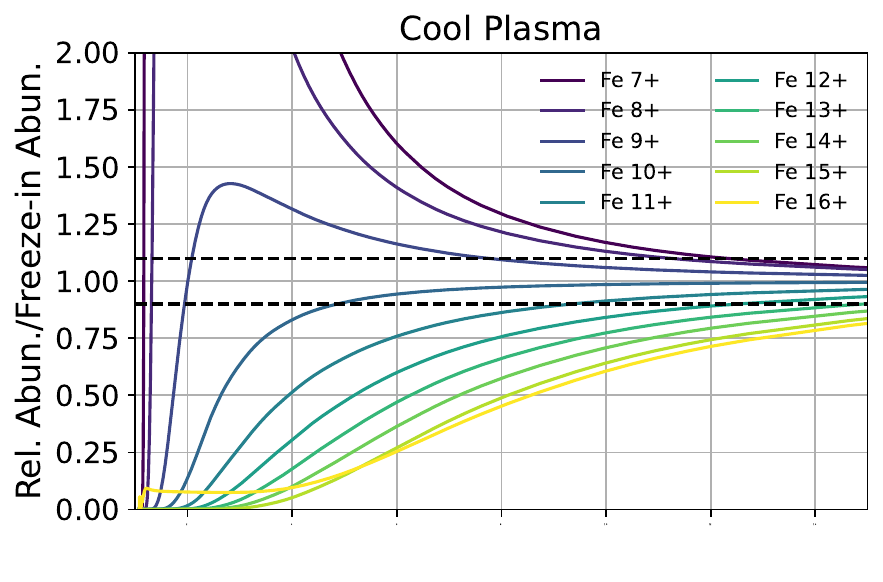}
\includegraphics[width=0.5\textwidth]{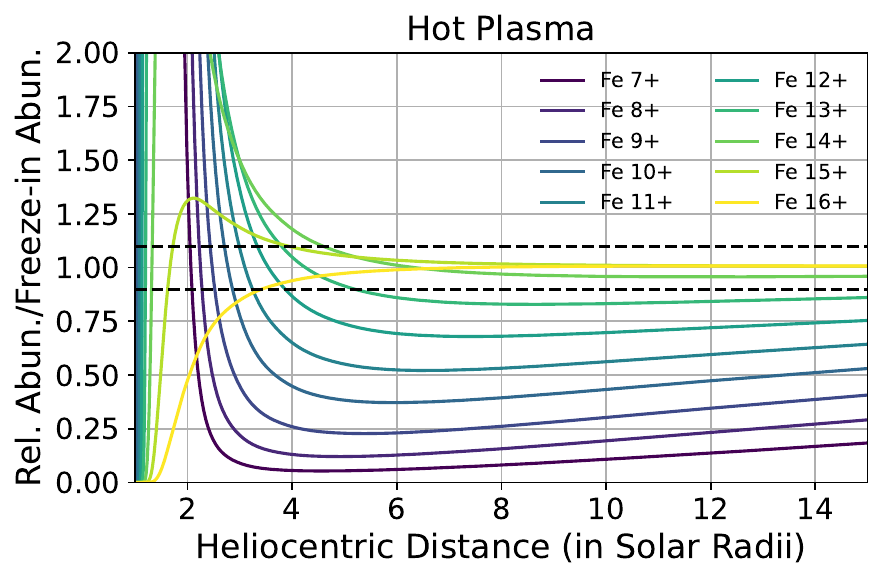}
\caption{\small Same as \ref{fig:freeze-in-plot_c}.}
\label{fig:freeze-in-plot_fe}
\end{figure}

Lastly, we visualize each species' charge-state evolution, plotting the charge states versus heliocentric distance of carbon, oxygen, and iron from the hot and cool prominence material in Figure \ref{fig:freeze-in-plot_c}, \ref{fig:freeze-in-plot_o}, \ref{fig:freeze-in-plot_fe}. The figures show two panels where the cool and hot plasmas' evolutions towards their freeze-in heights are shown in the top and bottom, respectively. The panels plot the relative abundances for individual ions normalized to their freeze-in values, or the value where the relative abundance no longer changes. The freeze-in distances are defined where the curves approach 1. The two horizontal dashed lines are at 0.9 and 1.1, indicating where their curves are within 10\% of their freeze-in altitude. For all three elements, the hot plasma's ions indicate a larger range of freeze-in distances compared to a much narrower span of altitudes from the cooler plasma. In particular, the hotter plasma shows two sets of freeze-in ranges where the most highly ionized ions  (e.g. C$^{4-6+}$) freeze-in close to the Sun compared to lower ionization states which are evolving farther away. However, as is shown in the distributions from Figure \ref{fig:best_fit_histogram}, the relative abundances of the lowest charge states simulated in the hot plasma have an insignificant overall contribution to the distribution itself. Overall, our simulated freeze-in distances range between 1.5 to $>15R_{\odot}$, in line with those found with 1D models of CMEs and prominences \citep{Gruesbeck2012, Rivera2019a} and 3D MHD models \citep{Rivera2023, Wraback2025}.

\section{Discussion}
\label{sec:discussion}

The work examines the radial evolution of a high latitude, complex, geoeffective quiescent filament eruption that was Earth-directed, measured both remotely and in situ. We used multi-viewpoint EUV measurements to determine the filament's thermodynamic evolution across the low to middle corona (Phase I). We then used these constraints as initial conditions and in situ measurements as final conditions for models of the thermodynamic evolution of the plasma from $1R_{\odot}$ out to 1au, and used these thermodynamic radial profiles to perform NEI modeling of the charge state distribution (Phase II).

Evaluating our Phase I results, we find that SDO/AIA and STEREO-A/EUVI indicate compatible derived temperature and density values. The temperature spans approximately $3-4\times10^5$ K while the number density varies over 2 orders of magnitude between 1$\times 10^{9} - 1\times 10^{11}$ \,cm$^{-3}$. The relative agreement across multiple, coordinated instruments lends credence to the methodology. The derived number density ranges between typical densities for prominences at the Sun, while the temperature reflect typical chromospheric temperature, often observed in pre-eruptive filaments as well \citep{Parenti2014}.  The fact that the central part of the prominence remained mostly in absorption across the EUV channels also supports the relatively cooler temperatures found. These results suggest that at least some part of the prominence plasma remained cool leaving the low corona and that at least part of it remained in a low ionized (or even neutral) state as it propagated into interplanetary space.

The mean column density values we found for AIA and STEREO-A were $1.5\times10^{10}$ $cm^{-3}$ and $3.1\times10^{9}$ $cm^{-3}$, respectively. There is roughly an order of magnitude difference between the two spacecrafts that we attribute to the imagers' different perspectives and line of sight (given they had a longitudinal separation of roughly 70 degrees). Although we generally expect the rapid expansion to progressively lower the density of the plasma, AIA's head-on view of the prominence meant that material intercepted was ultimately different from that observed from EUVI, and as such, pile up in the boxes being tracked potentially increased the density in certain cases.

The near-constant, cool temperature during the initial eruption suggests that the prominence experienced some initial heating that is counteracted by expansion, radiative loss, or thermal conduction, as found in \cite{Landi2010, Lee2017, Rivera2023, Sheoran2023}. However, the multi-thermal state reflected by the charge states observed in the heliosphere provides additional clues that suggest at least two likely scenarios. The first is that prominence experienced non-uniform heating after those initial remote constraints that ultimately generate the low and high ionized charge states later observed. Alternatively, the prominence could have been uniformly heated but part of the prominence cooled more rapidly from its expansion, from larger radiative losses, or through conduction.

The charge states measured in the heliosphere also suggests that at least some part of the prominence underwent strong heating during the eruption. As shown in Figure \ref{fig:best_fit_histogram}, the relative abundances of all three elements contain very low and extremely ionized ions found together that are rarely seen in the solar wind, i.e. Fe$^{16+}$, C$^{3+}$, O$^{4+}$. In particular, the double peaked iron distribution is a quintessential ion signature of CMEs that has been attributed to multi-thermal structures of prominences in the past \citep{Rivera2019a, Rivera2023}. By virtue of our analysis, our work follows the part of the prominence that remains in absorption, i.e. the coolest part. However, as mentioned in Section \ref{sec:observations}, we see that STEREO-B captures the perspective of the prominence that it is predominately in emission. This suggests that along with the cool component we analyze, there is a co-spatial hotter component that is quickly ionized or even existing in its pre-eruptive state \citep{Habbal2010_hotshrouds}. This falls in line with our Phase II NEI modeling results where the extended set of charge state distributions (especially in iron) could only be reconstructed through a combination of at least two plasmas with a distinct thermodynamic history, as shown in Figure \ref{fig:plasma_profiles}.    

The NEI modeling results predict a peak in electron temperature between $2-3R_{\odot}$, suggesting continuous heating of the prominence beyond our field of view. However, a key limitation to our work is that we do not have constraints to temperature beyond the low corona. Several ongoing heating mechanisms have been proposed: ohmic heating, wave heating, turbulence via the kink instability, reconnection, energetic particles, shock heating \citep{Murphy2011}. In general, it is found that wave heating would be ineffective \citep{Landi2010}. A small shock is observed in situ (at the start of CME 3 in Figure \ref{fig:Wind_data}, however it may be too small to drive significant heating within the CME body, as has been observed in more energetic events i.e. Bastille day CME \citep{Rivera2023}. Also, since there was no associated flare or visible plasma sheet structure, we do not anticipate strong ohmic heating \citep{Reeves2019}. In the context of in situ observations, the prominence is found within the flux rope structure, where the flux rope is identified by \citealt{Liu2012, Mostl2012}. The filament is embedded within much larger substructure that spans multiple, interacting CMEs. We speculate, that based on the complexity of the magnetic environment in which the filament is embedded in situ, one idea is that it can undergo significant and continuous small-scale reconnection after the ejection that drives localized heating.

\section{Conclusions}
This work capitalizes on a unique opportunity to constrain the thermodynamic evolution of a well-observed, quiescent prominence eruption that survives in a cool, low-ionized state, beyond the corona. The work models the temperature, density, and speed of plasma constrained by a multi-perspective view of its initial liftoff from the Sun as well as from the heavy ion measurements collected during its passage at L1. The plasma state (electron temperature and density) in the low corona is derived from remote observations where the prominence remains in absorption across several EUV channels. While its outflow speed is estimated from a stereoscopic view of the eruption. 

The work finds: 

\begin{enumerate}
    \item From EUV images, part of the quiescent filament is observed in absorption in the corona and remains in absorption throughout the eruption in the low corona, but part of it is also seen in emission, suggesting the prominence has an initial multi-thermal state.
    \item The prominence's multi-thermal state can be described by at least a two plasma system that includes a hot and cool component. The relatively hot and cool components form the highly ionized and low-ionized charge states, respectively, observed together at L1. 
    
    \item The presence of a much hotter plasma next to a cooler plasma suggests that the prominence is either undergoing non-uniform heating, or is all strongly heated with the 'Cool' plasma's heating being balanced by thermal conduction, radiation, or adiabatic expansion
    
    \item For the heating experienced, the models predict a peak in electron temperature within a few R$_{\odot}$ of the Sun in both the 'Cool' and 'Hot' plasma, suggesting the heating processes persists beyond the low corona.
    
\end{enumerate}

Remote EUV observations of the prominence evolution in the low corona suggested a consistent temperature along the structure in absorption with a surrounding more ionized structure. In situ heavy ion observations also reveal that prominence structure is multi-thermal. Leveraging both remote and in situ data was essential to placing constraints to the period in between the available but limited observations.

Although, we include constraints at the Sun and in the heliosphere of a CME, we still lack the continuous multi-thermal monitoring required to fully constrain the thermodynamic evolution up to the point where the plasma is being actively heating and accelerated from the Sun \citep{Rivera2019b, Wraback2024_predictofflimbobs, Rivera_Badman2025}. In particular, we lack global, simultaneous temperature and density diagnostics beyond the low corona for energy budget estimates. Proposed missions such as Extreme Ultraviolet Coronal mass ejection and Coronal Connectivity Observatory (ECCCO) would produce the necessary extended observations important to constrain heating, shock development, and propagation within several R$_{\odot}$ to refine our current estimates \citep{Reeves2024AGU}.

\acknowledgements
The authors thank Enrico Landi for insight on the methodology in the paper. C.A.G acknowledges support from NASA grant 80NSSC22K0225 and the NSF REU solar physics program at SAO, grant number AGS-2244112. K.K.R. acknowledges support from the NASA HSO Connect program, grant number 80NSSC20K1283. 
\\ \\
This work made use of the version 6.0.1 of the Astropy (\url{http://www.astropy.org}) a community-developed core Python package and an ecosystem of tools and resources for astronomy  \citep{Astropy2013, Astropy2018, Astropy2022} and version 5.0.0 of the SunPy \citep{Sunpy2020} open source software package. We also acknowledge the use of the Python packages: fiasco \citep{Barnes2024} version 0.2.3, Numpy \citep{Numpy2020} version 1.24.3, Scipy \citep{Scipy2020} version 1.10.1 and Matplotlib \citep{Matplotlib2007} version 3.7.1.


\end{document}